\documentclass[a4paper]{AO4ELT}  

\usepackage{microtype}
\usepackage{biblatex}
\usepackage{amsmath,amsfonts,amssymb}
\usepackage{graphicx}
\usepackage{pst-all} 
\usepackage[colorlinks=true, allcolors=blue]{hyperref}
\def\bac{\ref@jnl{Bull. astr. Inst. Czechosl.}}
\def\caa{\ref@jnl{Chinese Astron. Astrophys.}}
\def\cjaa{\ref@jnl{Chinese J. Astron. Astrophys.}}
\def\icarus{\ref@jnl{Icarus}}           
\def\jcap{\ref@jnl{J. Cosmology Astropart. Phys.}}
\def\jrasc{\ref@jnl{JRASC}}             
\def\memras{\ref@jnl{MmRAS}}            
\def\na{\ref@jnl{New A}}                
\def\nar{\ref@jnl{New A Rev.}}          
\def\pra{\ref@jnl{Phys.~Rev.~A}}        
\def\prb{\ref@jnl{Phys.~Rev.~B}}        
\def\prc{\ref@jnl{Phys.~Rev.~C}}        
\def\prd{\ref@jnl{Phys.~Rev.~D}}        
\def\pre{\ref@jnl{Phys.~Rev.~E}}        
\def\prl{\ref@jnl{Phys.~Rev.~Lett.}}    
\def\pasa{\ref@jnl{PASA}}               
\def\pasp{\ref@jnl{PASP}}               
\def\pasj{\ref@jnl{PASJ}}               
\def\rmxaa{\ref@jnl{Rev. Mexicana Astron. Astrofis.}}%
\def\qjras{\ref@jnl{QJRAS}}             
\def\skytel{\ref@jnl{S\&T}}             
\def\solphys{\ref@jnl{Sol.~Phys.}}      
\def\sovast{\ref@jnl{Soviet~Ast.}}      
\def\ssr{\ref@jnl{Space~Sci.~Rev.}}     
\def\zap{\ref@jnl{ZAp}}                 
\def\nat{\ref@jnl{Nature}}              
\def\iaucirc{\ref@jnl{IAU~Circ.}}       
\def\aplett{\ref@jnl{Astrophys.~Lett.}} 
\def\apspr{\ref@jnl{Astrophys.~Space~Phys.~Res.}}
\def\bain{\ref@jnl{Bull.~Astron.~Inst.~Netherlands}} 
\def\fcp{\ref@jnl{Fund.~Cosmic~Phys.}}  
\def\gca{\ref@jnl{Geochim.~Cosmochim.~Acta}}   
\def\grl{\ref@jnl{Geophys.~Res.~Lett.}} 
\def\jcp{\ref@jnl{J.~Chem.~Phys.}}      
\def\jgr{\ref@jnl{J.~Geophys.~Res.}}    
\def\jqsrt{\ref@jnl{J.~Quant.~Spec.~Radiat.~Transf.}}
\def\memsai{\ref@jnl{Mem.~Soc.~Astron.~Italiana}}
\def\nphysa{\ref@jnl{Nucl.~Phys.~A}}   
\def\physrep{\ref@jnl{Phys.~Rep.}}   
\def\physscr{\ref@jnl{Phys.~Scr}}   
\def\planss{\ref@jnl{Planet.~Space~Sci.}}   
\def\procspie{\ref@jnl{Proc.~SPIE}}   

\makeatletter         
\def\@maketitle{
\includegraphics[width = 170mm]{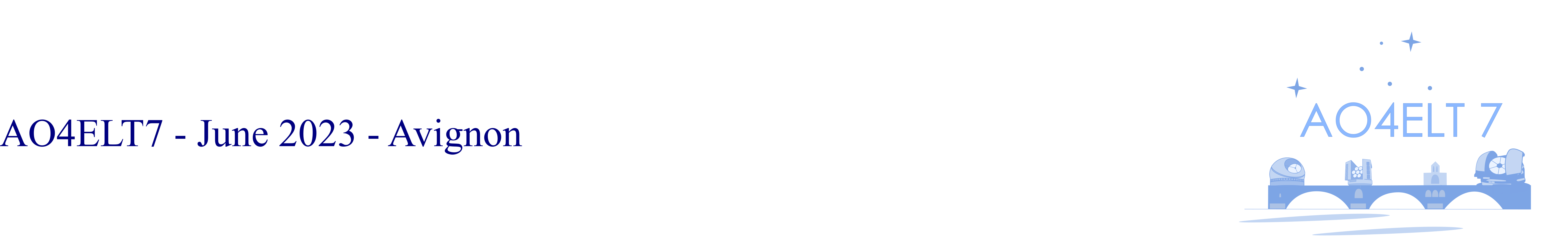}\\[8ex]
\begin{center}
{\Huge \bfseries \sffamily \@title }\\[4ex] 
{\Large  \@author}\\[4ex] 
\@date
\end{center}}

\title{SOUL at LBT: commissioning results, science and future.}

\author[a,b]{Enrico Pinna}
\author[a,b]{Fabio Rossi}
\author[a,b]{Guido Agapito}
\author[a,b]{Alfio Puglisi}
\author[a,b]{Cédric Plantet}
\author[a,b]{Essna Ghose}
\author[c]{Matthieu Bec}
\author[a,b]{Marco Bonaglia}
\author[a,b]{Runa Briguglio}
\author[c]{Guido Brusa}
\author[a,b]{Luca Carbonaro}
\author[c]{Alessandro Cavallaro}
\author[c]{Julian Christou}
\author[d]{Olivier Durney}
\author[d,c]{Steve Ertel}
\author[a,b]{Simone Esposito}
\author[a,b]{Paolo Grani}
\author[c]{Juan Carlos Guerra}
\author[d]{Philip Hinz}
\author[c]{Michael Lefebvre}
\author[a,b]{Tommaso Mazzoni}
\author[c]{Brandon Mechtley}
\author[c]{Douglas L. Miller}
\author[d]{Manny Montoya}
\author[c]{Jennifer Power}
\author[c]{Barry Rothberg}
\author[c]{Gregory Taylor}
\author[d]{Amali Vaz}
\author[a,b]{Marco Xompero}
\author[c]{Xianyu Zhang}

\affil[a]{INAF -- Osservatorio Astrofisico di Arcetri, Largo E. Fermi 5, 50125, Firenze, Italy}
\affil[b]{ADONI - ADaptive Optics National laboratory in Italy}
\affil[c]{Large Binocular Telescope Observatory, The University of Arizona, 933 North Cherry Ave, Tucson, AZ 85721, USA}
\affil[d]{Department of Astronomy and Steward Observatory, The University of Arizona, 933 North Cherry Ave, Tucson, AZ 85721, USA}
\authorinfo{Further author information: (Send correspondence to Enrico Pinna)\\E-mail: enrico.pinna@inaf.it}

\begin{document} 
\maketitle

\begin{abstract}
The SOUL systems at the Large Bincoular Telescope can be seen such as precursor for the ELT SCAO systems, combining together key technologies such as EMCCD, Pyramid WFS and adaptive telescopes. After the first light of the first upgraded system on September 2018, going through COVID and technical stops, we now have all the 4 systems working on-sky. Here, we report about some key control improvements and the system performance characterized during the commissioning. The upgrade allows us to correct more modes (500) in the bright end and increases the sky coverage providing $SR(K)>20\%$ with reference stars $G_{RP}<17$, opening to extragalcatic targets with NGS systems. Finally, we review the first astrophysical results, looking forward to the next generation instruments (SHARK-NIR, SHARK-Vis and iLocater), to be fed by the SOUL AO correction.

\end{abstract}


\section{INTRODUCTION}
\label{sec:intro}  
\begin{figure}[ht]
    \centering
    \includegraphics[width=0.85\linewidth]{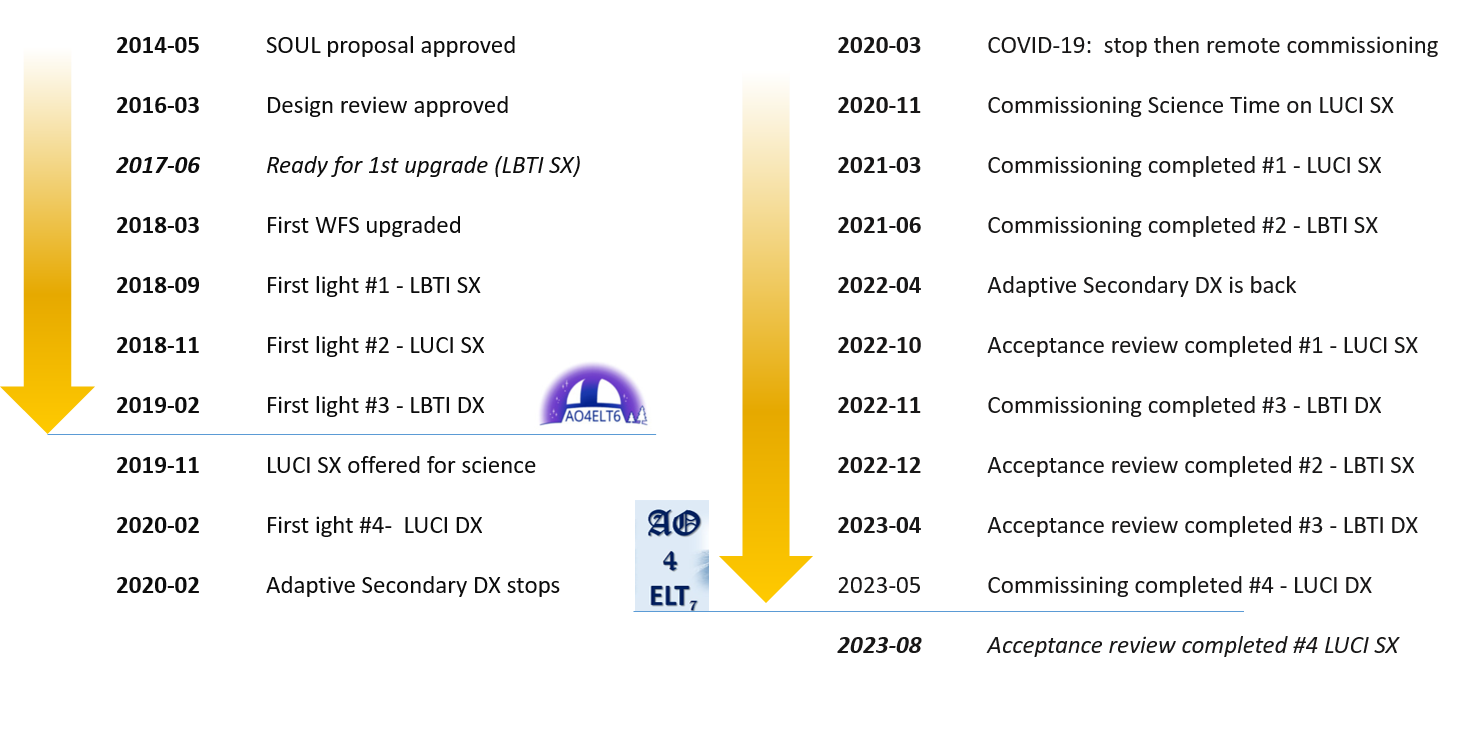}
    \caption{The temporal evolution of the project from its proposal to the acceptance of the AO systems. The different systems are identified with the instrument name (LUCI/LBTI) and the binocular telescope side (SX/DX).}
    \label{fig:timeline}
\end{figure}

The SOUL project consists of the upgrade of the 4 Single Conjugate Adaptive Optics (SCAO) Natural Guide Star (NGS) systems of the Large Binocular Telescope. In fig.~\ref{fig:timeline}, We report the main steps of the project, starting from the approval of the proposal in 2014 until today, when the last of the four systems has been formally accepted. The upgrade itself and its first results have been previously reported in Refs. \cite{Pinna_SOUL_SPIE_2016} and \cite{SOUL_on_sky_AO4ELT6_Pinna2019} respectively. Here, we just recall the main goal of the upgrade: exploiting the Electron-Multiplying Charge Coupled Detector (EMCCD) technology by replacing the CCD39 (Scimeasure) with Ocam2k (First Light Imaging) cameras as detectors \cite{EMCCD_Agapito_2018} \cite{Pinna_SOUL_SPIE_2016} for the AO system. The upgrade eventually has a wide impact in the HW and SW of the original systems \cite{Esposito_FLAO_SPIE2010} \cite{Esposito_FLAO_SPIE2011}, enabling: higher frame rate and pupil sampling and a more sophisticated control (see sect.~\ref{sec:key_steps}).

The 4 SCAO systems are distributed in pairs on the two sides (SX and DX) of the LBT. On each eye of the telescope we have one Adaptive Secondary Mirror (ASM) \cite{Riccardi_ASM_SPIE2010} \cite{Brusa_SPIE2020} and two focal stations fed by SOUL: LUCI (NIR spectro-imager) \cite{Seifert_LUCI_SPIE2003} and LBTI (interferometer) \cite{Hinz_LBTI_SPIE206} . 
The 2 LUCI instruments work independently, while the 2 LBTI foci can be combined for the interferometric modes. In 2018, when the first system was upgraded, all the 4 SCAO were offered to the community for scientific observations. The upgrade strategy was aimed to minimize the impact on scientific observations, while providing the benefits of the SOUL upgrade to the 4 focal stations in the shortest time. This resulted in a parallel commissioning on the 4 systems, divided in two main steps, as reported in fig.~\ref{fig:timeline}. The first one includes the WFS and SW upgrades, the WFS installations and its alignment at the focal station, a preliminary calibration completed by the first light of the AO system. At this stage, the system is delivering AO performances as good as or better than FLAO systems and is offered for science observations. The results reported at AO4ELT6 \cite{SOUL_on_sky_AO4ELT6_Pinna2019} refer to this first stage for the 2 SX systems. The second phase consists in the full calibration, the parameter tuning and on-sky characterization of the performance. The end of this phase corresponds to the commissioning completion. This phase is then followed by a formal acceptance review process held by LBTO. In August 2023 all the 4 systems have been formally accepted by the telescope. 

In this paper, we report a set of information collected in the commissioning experience that we hope can be useful to the community for the design and development of the next generation of SCAO systems. In the next section we select 3 control improvements, in sect.~\ref{sec:performance} we describe the performance of the last commissioned system and, finally, in sect.~\ref{sec:science} we report the early science results.

\section{Key control improvements}
\label{sec:key_steps}

Here, we highlight three key control advancements identified during the SOUL commissioning activity, which have had the most significant impact on system performance and reliability. All of them are now routinely used during science operations on all the systems. 
These three improvements are:
\begin{itemize}
    \item IIR Control for tip/tilt modes at high framerate
    \item Leaky integrator for higher modes
    \item Pyramid optical gain compensation
\end{itemize}

\begin{figure}[ht]
    \centering
    \includegraphics[width=0.5\linewidth]{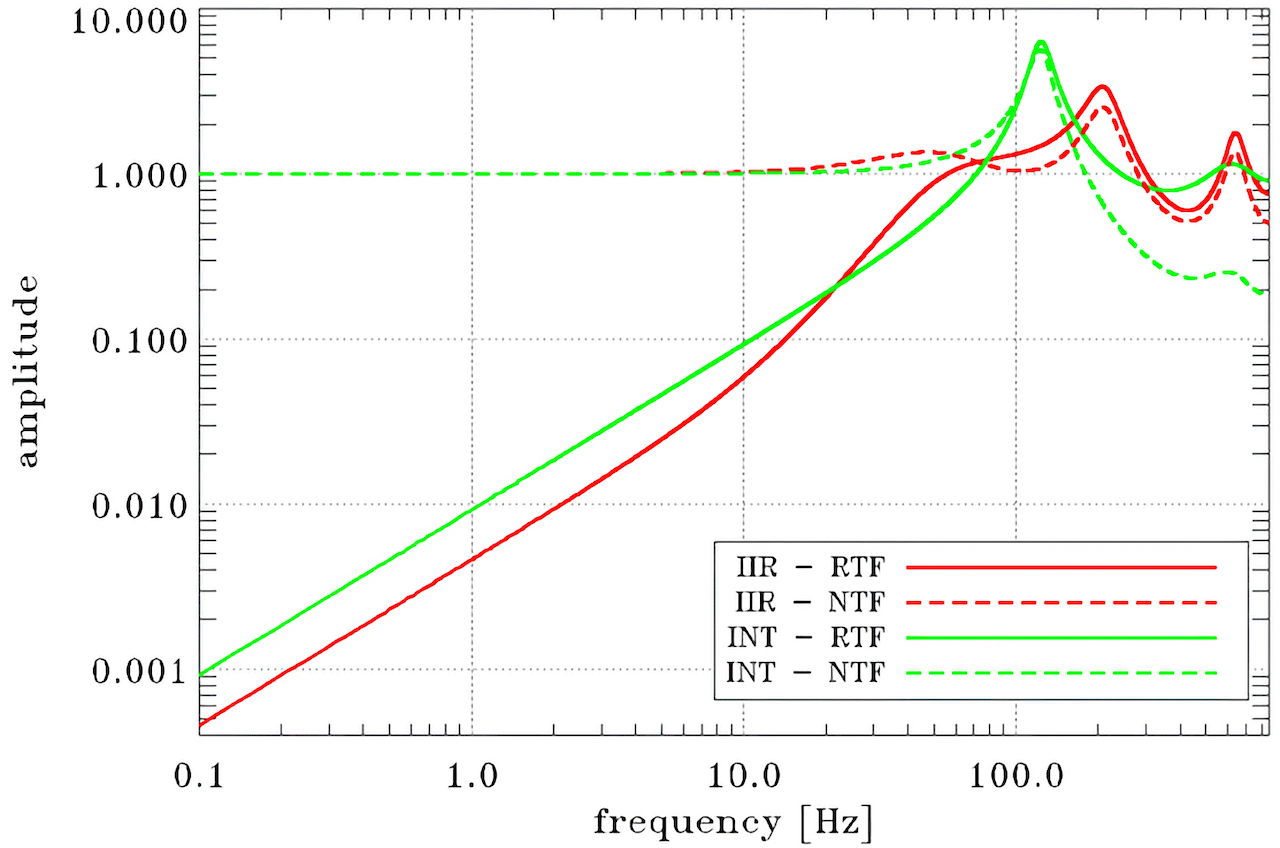}
    \caption{The rejection (solid) and noise (dashed) transfer functions for a  pure integrator (green) and IIF (red) temporal control, considering the SOUL AO loop delay and a framerate of $1.7$kHZ. The IIF filter improves the rejection at low frequencies, while keeping the noise overshoot out of the critical frequency of $\sim120$Hz.}
    \label{fig:IIR_TF}
\end{figure}

The AO temporal control in the FLAO systems \cite{Esposito_FLAO_SPIE2010} has been a modal pure integrator, delivering the expected performance. The SOUL upgrade planned to keep the same kind of control.
However, SOUL pushed the AO loop frame rate up to $1.7kHz$ and reduced the delay down to 2.0ms, thanks to Ocam2k.
Raising the gain, in order to reach the expected rejection around $13Hz$\footnote{Around $13Hz$ we have the resonance of the telescope swing arms holding ASM and M3 \cite{2012SPIE.8447E..1CK}.
During on-sky operations, the lines around this frequency are the major contributors of the vibration spectrum.}, the noise transfer function of the SOUL pure integrator amplified the noise around $120Hz$, as shown in green in fig.~\ref{fig:IIR_TF}. 
During the commissioning operations, we discovered that the adaptive secondary mirror of LBT presented a resonant frequency in this range.
This resonance is excited by the pure integrator control when the gain of tip-tilt modes is increased in order to provide the expected rejection at frequencies $10-20Hz$.
The identified solution has been the implementation of Infinite Impulse Response (IIR) temporal filters for tip and tilt modes.
This approach allows us to shape the transfer functions improving the rejection at $10-20Hz$, while pushing the noise amplification out of the resonance frequency, as shown by the red curves in fig.~\ref{fig:IIR_TF}.
The detailed description of the implementation and the test on the AO systems can be found in Ref.~\cite{Agapito_Control_MNRAS2021}.
Here, we want to mention that IIR filters is now implemented on the 4 SOUL systems and is active on tip-tilt modes for AO loop frame rates higher than $600Hz$, allowing us to meet the expected AO correction, see sect.~\ref{sec:performance}.

\begin{figure}[ht]
    \centering
    \includegraphics[width=0.4\linewidth]{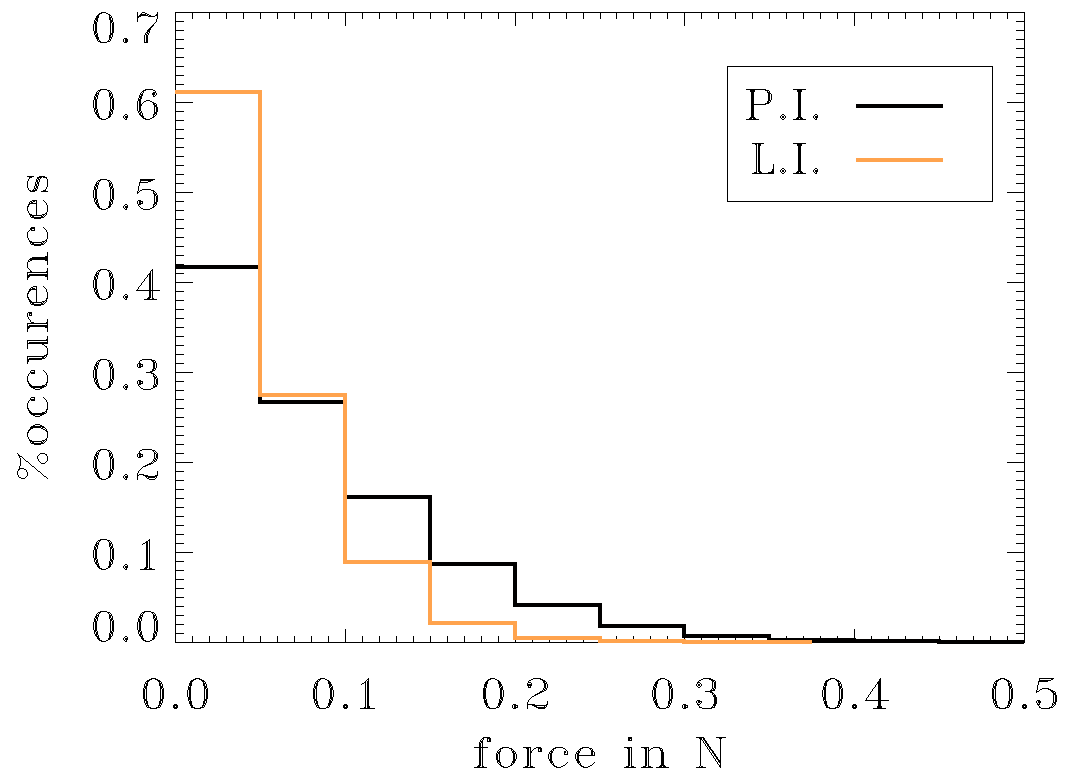}
    \caption{The distribution of the ASM actuator forces, during closed loop operations on-sky. We compare the cases of pure integrator (black) and a leaky integrator (orange) for the AO temporal control on higher modes. Both cases are measured under the same conditions.}
    \label{fig:Leak_force}
\end{figure}

The second control improvement we want to mention here is the adoption of the leaky integrator for higher modes. As proposed and demonstrated at AO4ELT6 \cite{Agapito_Goldfish_AO4ELT6_2019}, the introduction of a ``forgetting'' factor into the pure integrator control scheme of higher controlled modes, is beneficial for AO operations. The leakage allows us to reduce the impact of high order static commands coming from residual misalignment between WFS and corrector or those that are present in the initial shape of the mirror. These effects introduce a small wavefront error, but, because of their high spatial scale, it requires a relevant inter-actuator stroke, affecting the force budget of the adaptive secondary mirror. The leaky integrator has a low benefit on wavefront error, but a high return in terms of used forces. In fig.~\ref{fig:Leak_force}, we report one of the results presented in Ref.~\cite{Agapito_Goldfish_AO4ELT6_2019}, showing an example of gain in force distribution obtained by the leaky integrator versus the pure integrator. The leaky integrator is now implemented and active on the SOUL systems at any configuration (loop framerate and number of corrected modes). This allows SOUL to operate even beyond the nominal limit of $1.5asec$ of seeing.

\begin{figure}[ht]
    \centering
    \includegraphics[width=0.49\linewidth]{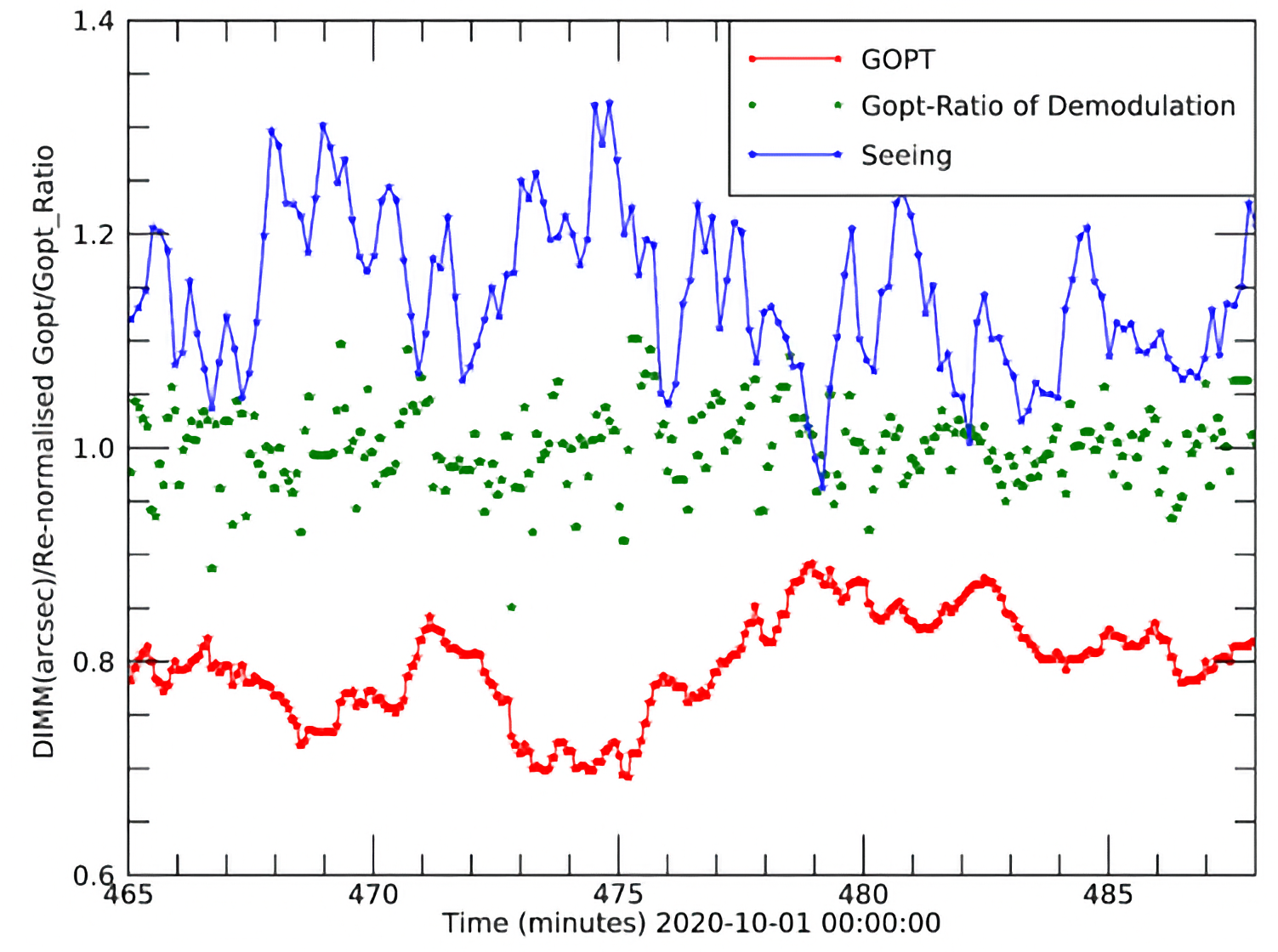}
    \includegraphics[width=0.49\linewidth]{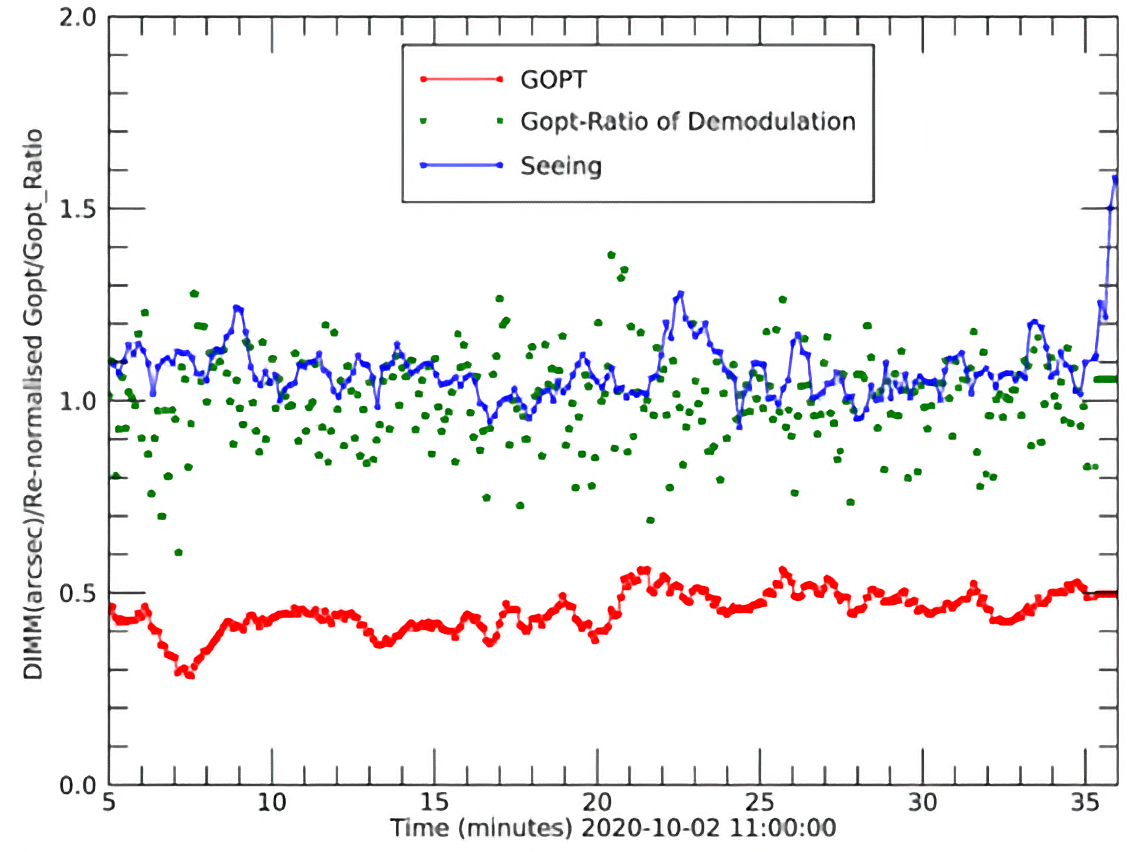}
    \caption{Measurement and correction of the PWFS optical gain during on-sky operations. The plots report the value of the optical gain compensated in the WFS measurements (red), the ratio between applied and measured probe signal (green) and the seeing, as measured by the DIMM (blue). The left plot is with AO reference star is $R\sim10$, while on the right is $R\sim16$.}
    \label{fig:Gopt}
\end{figure}

The third control improvement we report is the measurement and compensation of the Pyramid WaveFront Sensor (PWFS) optical gain. As well known from the theory \cite{2001_Esposito_PWFS-partialcorr_AA}, the pyramid WFS changes its sensitivity depending on the amplitude of the residual wavefront aberrations.
In order to apply the desired correction of non-common-path aberrations and keep the same AO loop temporal gain under the natural evolution of the seeing, it is mandatory to measure and compensate the variation of PWFS optical gain. This has been already detailed at AO4ELT5 \cite{Esposito_Gopt_AO4ELT5_2015}, together with the first proposed technique and demonstration of measurement and correction of the optical gain.
In brief, the technique is based on the injection of a calibrated probe signal on the adaptive secondary and its measurement on the PWFS during operations.
Details about this technique and first on-sky results with FLAO at LBT are reported in Ref.~\cite{Esposito_Gopt_A&A_2020}. 
In literature, other techniques have been proposed, as for example Refs. \cite{2019Deo_PWFS_gopt_modal_A&A...629A.107D} and \cite{2020Chambour_PWFS_Gopt_convol_A&A...644A...6C}. 
During the SOUL commissioning we worked more on the tuning of the technique proosed in \cite{Esposito_Gopt_A&A_2020} with the aim of improving its temporal response, following the seeing evolution and enabling its use even with AO reference source in the faint end of the magnitude range. We report in fig.~\ref{fig:Gopt} two examples of optical gain compensation on-sky on bright ($R\sim10$ on the left) and faint ($R\sim16$ right) AO references. The red line is the measurement of the PWFS optical gain, while the green dots report the ratio between the injected and measured amplitude of the probe signal. A ratio of $1$  means a perfect compensation of the PWFS optical gain on the probe mode. The bright case shows an highly variable seeing (blue dots are the differential image motion monitor -- DIMM -- measurement along the line of sight) in the range $0.95-1.30"$. Despite this variability the optical gain compensation is able to keep the ratio value of 1.0 with $\sigma=0.04$ and P2V of 0.2. The deviation of ratio from 1.0 is a direct estimation of the error we introduce in the compensation of the optical gain on the probe mode. In the faint case, the seeing is more stable, while the AO residuals are higher with an optical gain reduced down to $0.4-0.5$. Under these conditions the optical gain compensation keeps the ratio value at 1.0 with $\sigma=0.11$. 

\begin{figure}[ht]
    \centering
    \includegraphics[width=0.49\linewidth]{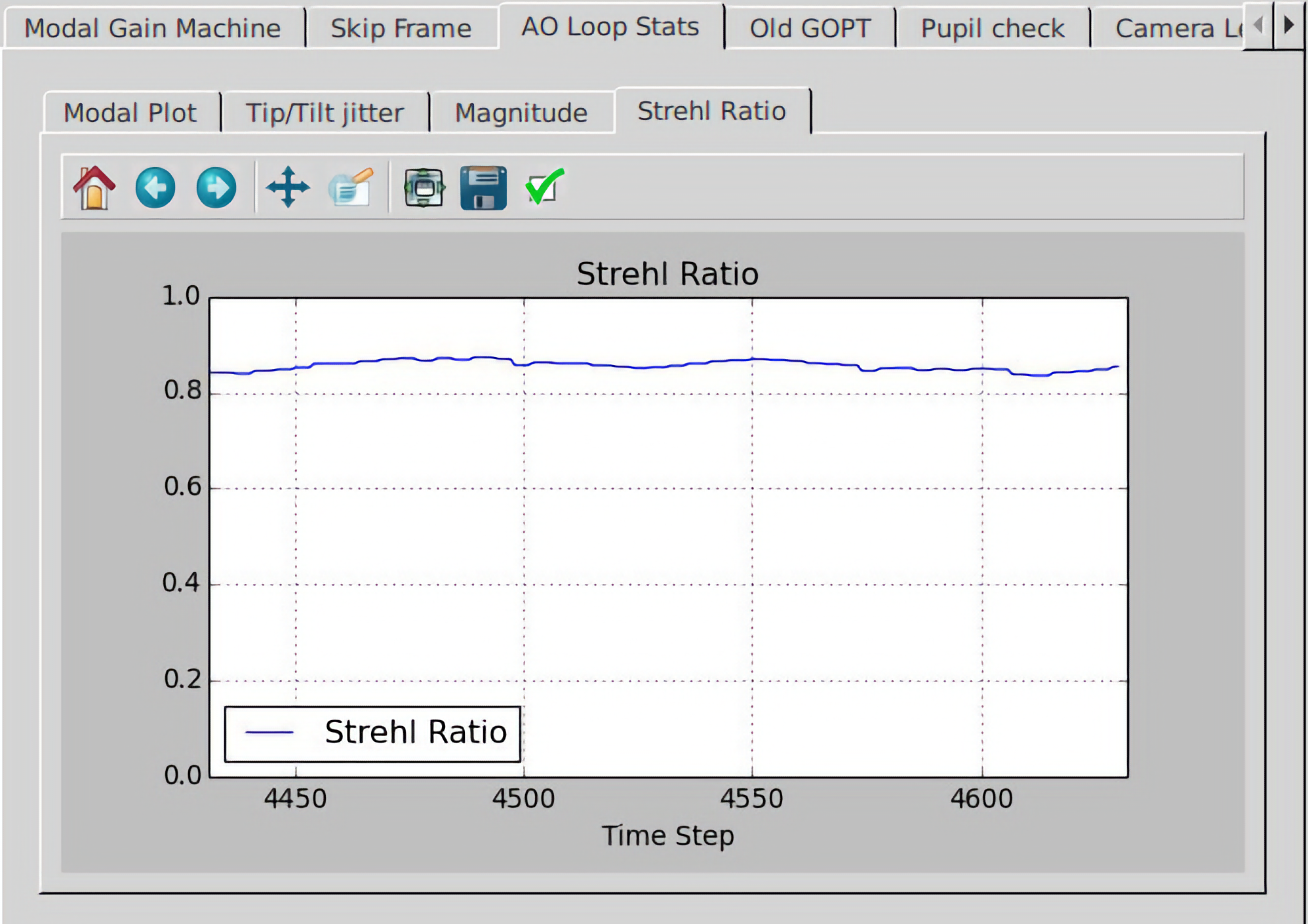}
    \includegraphics[width=0.49\linewidth]{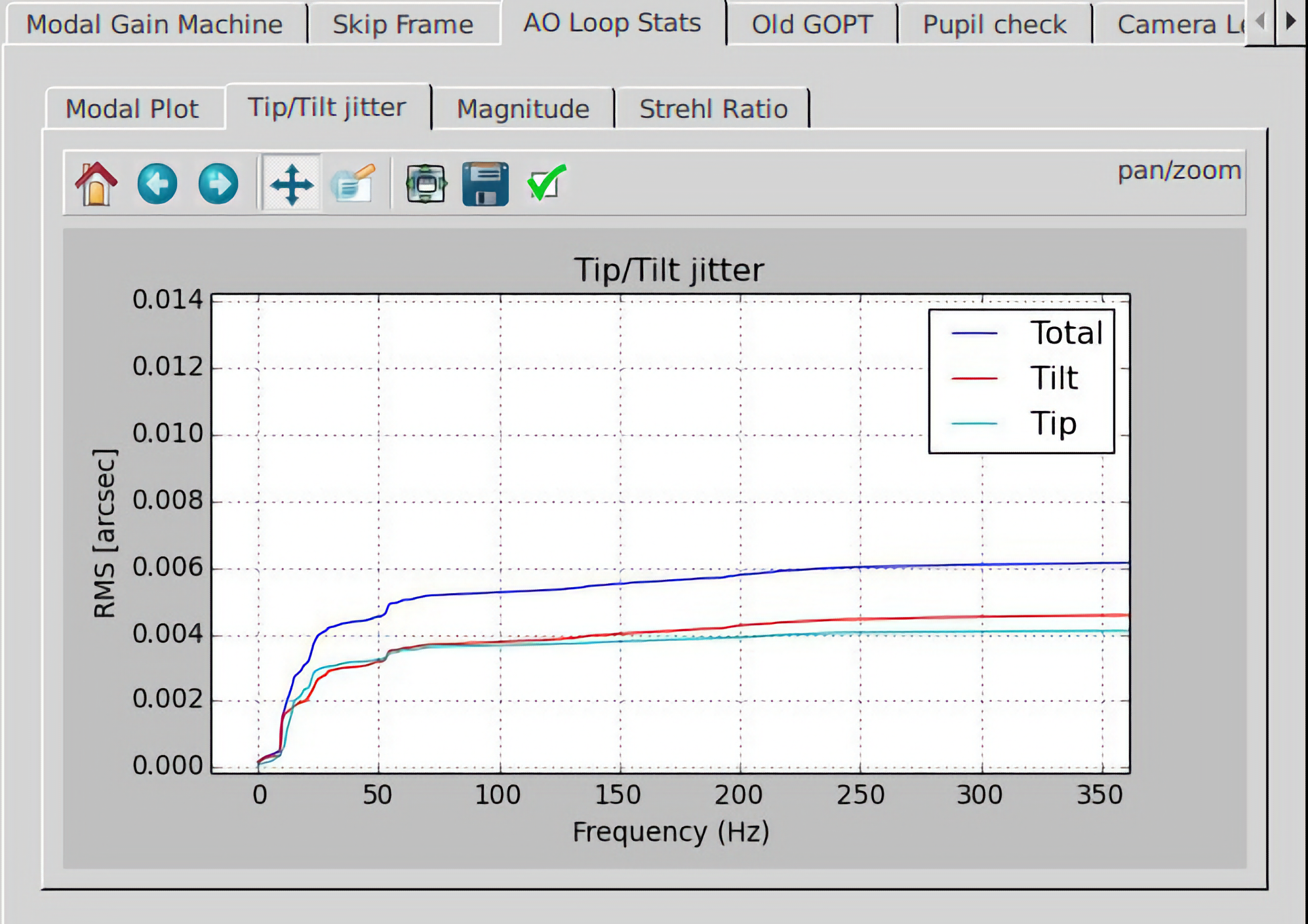}
    \caption{Example of 2 diagnostic windows available for the operator during AO operations. On the left, the SR(H) value estimated by the PWFS residuals versus time.
    On the right, the temporal cumulative spectrum of residual tip, tilt and their quadratic sum. These plots are refreshed at about $\sim0.5Hz$. 
    Many other diagnostic plots are available to the operator on the window tabs, as the residual WF versus mode number and optical gain value.}
    \label{fig:Auxloop_plot}
\end{figure}

The correction of the optical gain, together with the main advantages mentioned above, provides a continuous calibration of the AO residuals, as measured by the WFS. This means to have a real time estimation of the AO residuals that is translated by dedicated tools into Strehl Ratio (SR) and residual Point Spread Function (PSF) jitter. These values are made available to the operators who can monitor in real time the correction behaviour during science operations. We report in fig.~\ref{fig:Auxloop_plot} a couple of examples of the windows available to the operators with the plot of the SR at 1650nm and the residual PSF jitter as obtained from the WFS measurements compensated for the current PWFS optical gain. A larger set of data is then saved at around $1Hz$ in a database for the offline check of the AO performance for debug and support of scientific data reduction (an example of the data extracted from the database is shown in fig.~\ref{fig:Gopt}).  

\section{AO performance}
\label{sec:performance}

In this section, we report the performance of SOUL, as measured on the SOUL-LUCI-DX system. This is the last of the 4 that have been commissioned, taking advantage of all the improvements gathered during the commissioning activities that began in 2018.\footnote{We want to mention here that the SOUL software is unique for the 4 systems. It is running on different machines for the different systems: one workstation for each of the 4 WFSs and one for each of the 2 adaptive secondaries, but keeping an unique git repository. Of course, some of the configurations differ from system to system, but we made a constant effort to keep the configuration differences as minimal as possible. This strategy is aimed to ease the porting of any update from one system to the others with great benefit for the maintenance. In brief, the performance reported here are measured on a single system, but can be considered as representative of all the 4 systems.} 

\begin{figure}[ht]
    \centering
    \includegraphics[width=0.48\linewidth]{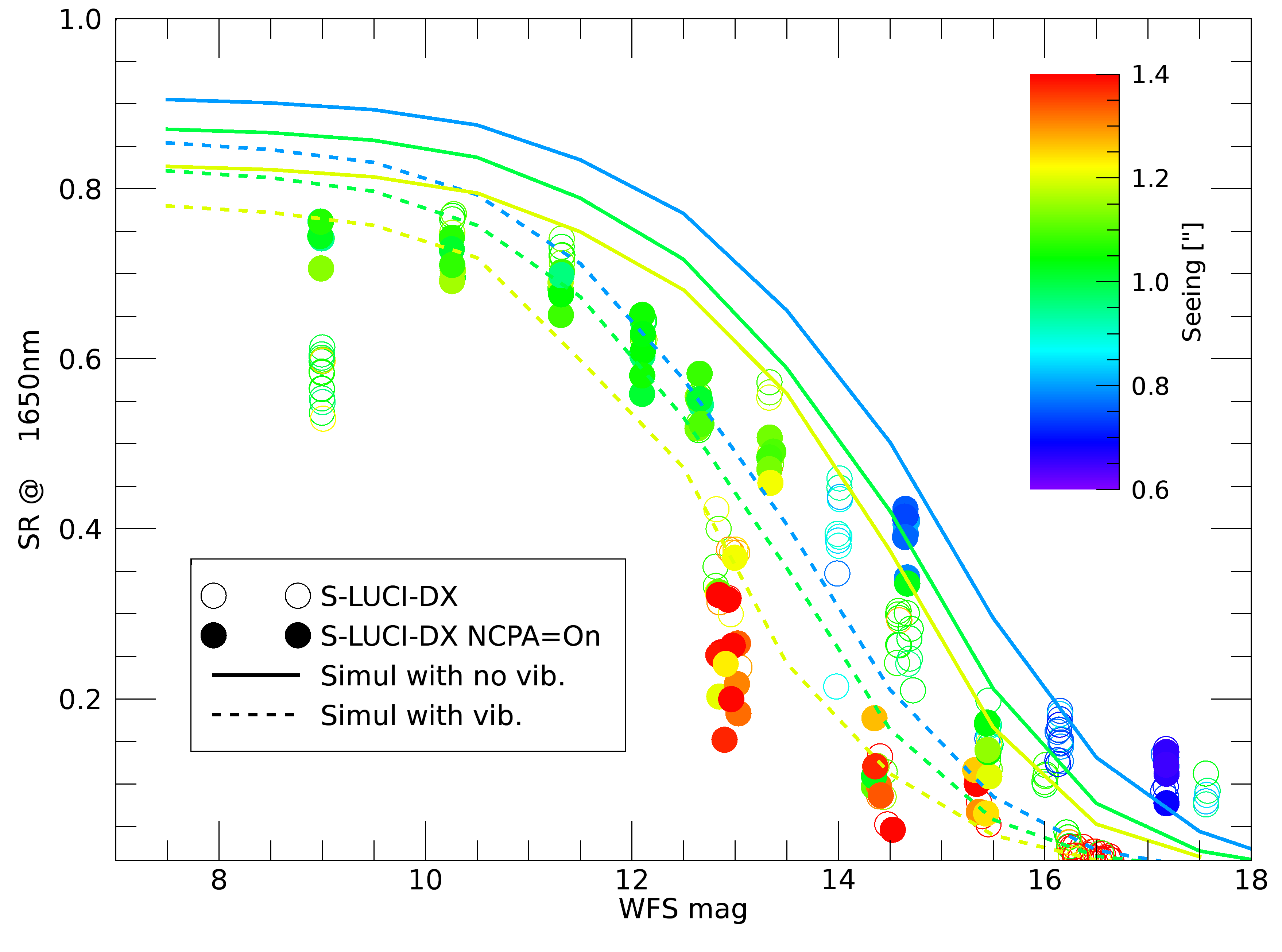}
    \includegraphics[width=0.49\linewidth]{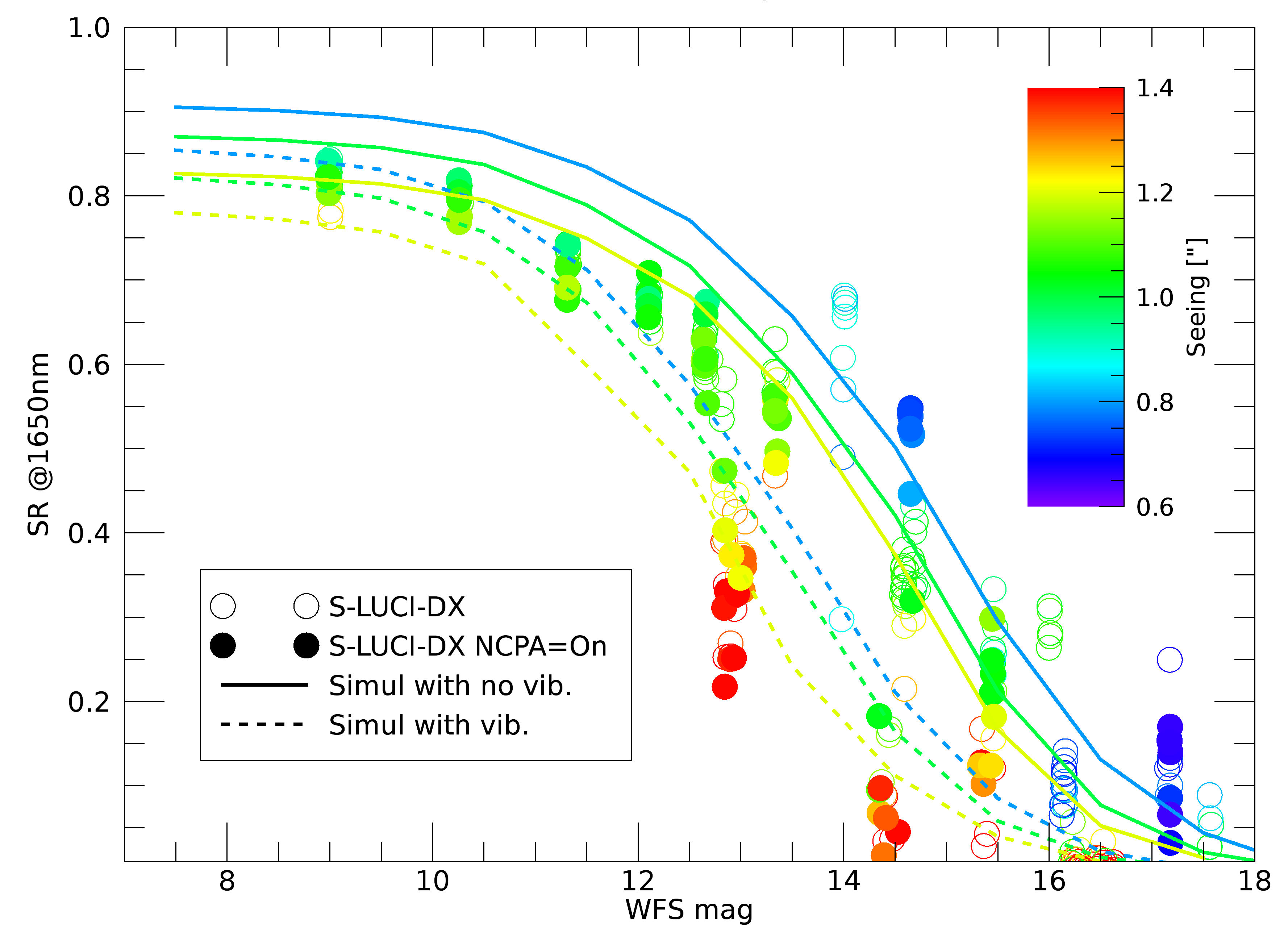}
    \caption{on-sky performance of SOUL-LUCI-DX, as measured during its commissioning (points) compared with numerical simulations (lines). Left: SR as measured on the LUCI2 N30 camera and scaled to 1650nm.
    Right: the same data-sets represented on the left, with SR estimated via SOUL telemetry from the AO residuals.
    The color of the points represents the seeing along the line of sight, as measured by the DIMM. On the horizontal axis, we report the AO reference star brightness expressed in WFS magnitude, as detailed in the text.}
    \label{fig:SR_SKY}
\end{figure}

In fig.~\ref{fig:SR_SKY} we plot the SR values measured on-sky during the commissioning, as estimated from the PSF on LUCI2 N30 camera (left) and from the SOUL fast telemetry (right). The PSF data have been measured on the AO reference star imaged by the LUCI2 N30 camera \cite{Seifert_LUCI_SPIE2003}, using different filters depending on the star brightness: FeII for the brightest ones, H for the central range of magnitudes and K for the fainter ones. The PSF SR plot has been produced by scaling, with Maréchal's approximation, the SR values to $1650nm$ from the central wavelength of the filter used during the image acquisition.

On the horizontal axis of the plots in fig.~\ref{fig:SR_SKY}, we report the ``WFS mag'' that is a magnitude value computed from the AO reference flux, as measured by the WFS camera.
This quantity expresses the amount of photons available for wavefront sensing.
The zero of the WFS mag has been arbitrarily defined as the one used for the numerical simulations \cite{SOUL_on_sky_AO4ELT6_Pinna2019} \cite{Agapito_semiAnalyt_MNRAS2019}.
This choice eases the comparison (fig.~\ref{fig:SR_SKY}) between the expected performances (solid lines for no vibrations and dotted lines for high vibrations) and those measured on-sky (points).
In the plots of fig.~\ref{fig:SR_SKY}, we can see a good agreement between the simulations and the measurements with the only exception being the PSF SR in the bright regime.
This is due to the high level of non-common-path aberrations (peak $\sim150nm$ RMS of wavefront \cite{Esposito_Gopt_A&A_2020}) of the N30 camera in LUCI2.
This aberration varies with the instrument rotation angle and time drifts limits the accuracy of their characterization. The resulting effect is that the NCPA compensation can recover a good optical quality, but it does not allow to reach values above $80\%$ at $1650nm$.    

The SOUL systems, as previously the FLAO ones, changes its configuration depending on the flux provided by the AO reference.
As described in Ref. \cite{SOUL_on_sky_AO4ELT6_Pinna2019}, the WFS mag is the number on which the system configuration is based.
In tab.~\ref{tab:Tabellone}, we report the main configuration parameters as refined during the SOUL-LUCI-DX commissioning. During the AO setup, the system measures the actual WFS mag of the reference star; then, it configures the AO parameters, interpolating the value of the table. 
The reported parameters are the WFS camera binning, the AO loop framerate, the electron-multiplying (EM) gain of the WFS camera and its binning.
The camera binning sets the pupil sampling to 40, 20, 13 ad 10 sub-apertures (SA) on diameter for 1$\times$1, 2$\times$2, 3$\times$3 and 4$\times$4, respectively. The binning is done on chip in order to minimize the read out noise. First light Imaging developed custom read out modes on Ocam2k for SOUL, as the binning 3$\times$3 and the cropped 1$\times$1 (no binning). This last mode allows to read only 120 lines (those in the SOUL region of interest), providing a read out time of $240\mu s$. 

\begin{table}[ht]
\caption{The main AO parameters for SOUL-LUCI-DX. The automatic configuration of the AO system is based on the star magnitude, as measured on the AO WFS (WFS mag).} 

\label{tab:Tabellone}
\begin{center}       
\begin{tabular}{|c|c|c|c|} 
\rule[-1ex]{0pt}{3.5ex}  \textbf{WFS mag} & \textbf{Binning} & \textbf{Framerate [kHz]} &  \textbf{EM gain}\\
\hline
2.5 & 1$\times$1  & 1.70  & 4  \\
3.5 & 1$\times$1  & 1.70  & 10 \\
4.5 & 1$\times$1  & 1.70  & 25 \\
 5.5 & 1$\times$1  & 1.70  & 64 \\
  6.5 & 1$\times$1  & 1.70  & 163 \\
 7.5 & 1$\times$1  & 1.70  & 410 \\
 8.5 & 1$\times$1  & 1.70  & 600 \\
 10.5 & 1$\times$1  & 1.70  & 600 \\
 11.4 & 1$\times$1  & 1.35  & 600 \\
11.7 & 1$\times$1  & 1.00  & 600 \\
12.5 & 1$\times$1  & 0.75  & 600 \\
 12.7 & 2$\times$2  & 1.25  & 600 \\
 13.0 & 2$\times$2  & 1.00  & 600 \\
 14.0 & 2$\times$2  & 0.80  & 600 \\
14.5 & 3$\times$3  & 1.00  & 600 \\
 15.0 & 3$\times$3  & 0.75  & 600 \\
16.5 & 4$\times$4  & 0.25  & 600 \\
 17.5 & 4$\times$4  & 0.17  & 600 \\
\end{tabular}
\end{center}
\end{table}

\begin{figure}[ht]
    \centering
    \includegraphics[width=0.65\linewidth]{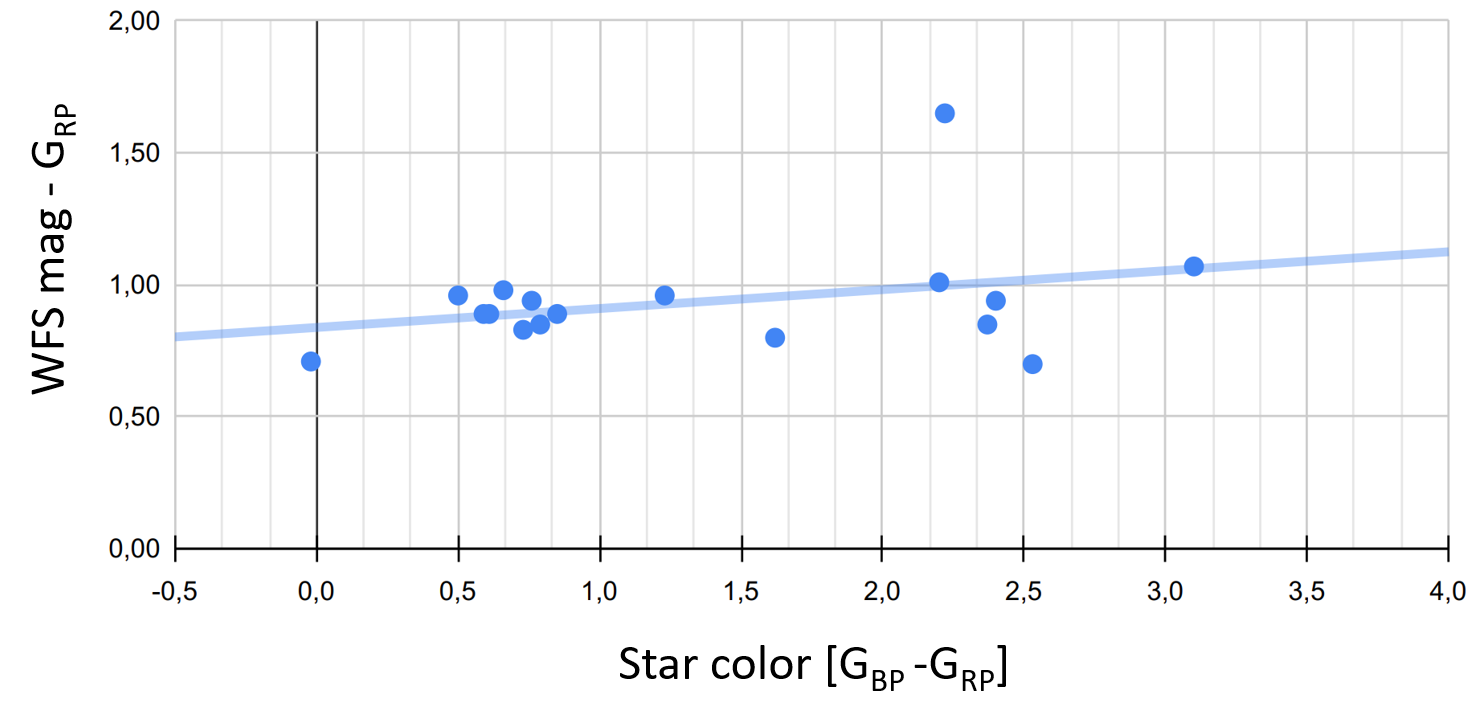}
    \caption{Calibration of the WFS mag with respect to GAIA $G_{RP}$. On the horizontal axis we report the star color. The plot shows that we can consider $WFS-mag = G_{RP}+0.9$, neglecting a mild color dependency.}
    \label{fig:Gaia_color_mag}
\end{figure}

On one hand, the WFS mag is useful for comparing the performance of the system with the numerical simulations, whereas on the other hand, it needs to be calibrated to be converted to catalogue magnitudes.
The initial calibration was done in with R magnitude, but there was a strong color dependency due to the WFS wavelength band ($600-950nm$), extending widely in I band. With the public access of GAIA catalogue, we decided to use $G_{RP}$ (GAIA RP magnitudes) as preferred catalogue magnitude, since its band coverage is very close to the WFS one.
In fig.~\ref{fig:Gaia_color_mag}, we can see a calibration set, with very low color dependency between WFS and $G_{RP}$ mags. 

\begin{figure}[ht]
    \centering
    \includegraphics[width=0.7\linewidth]{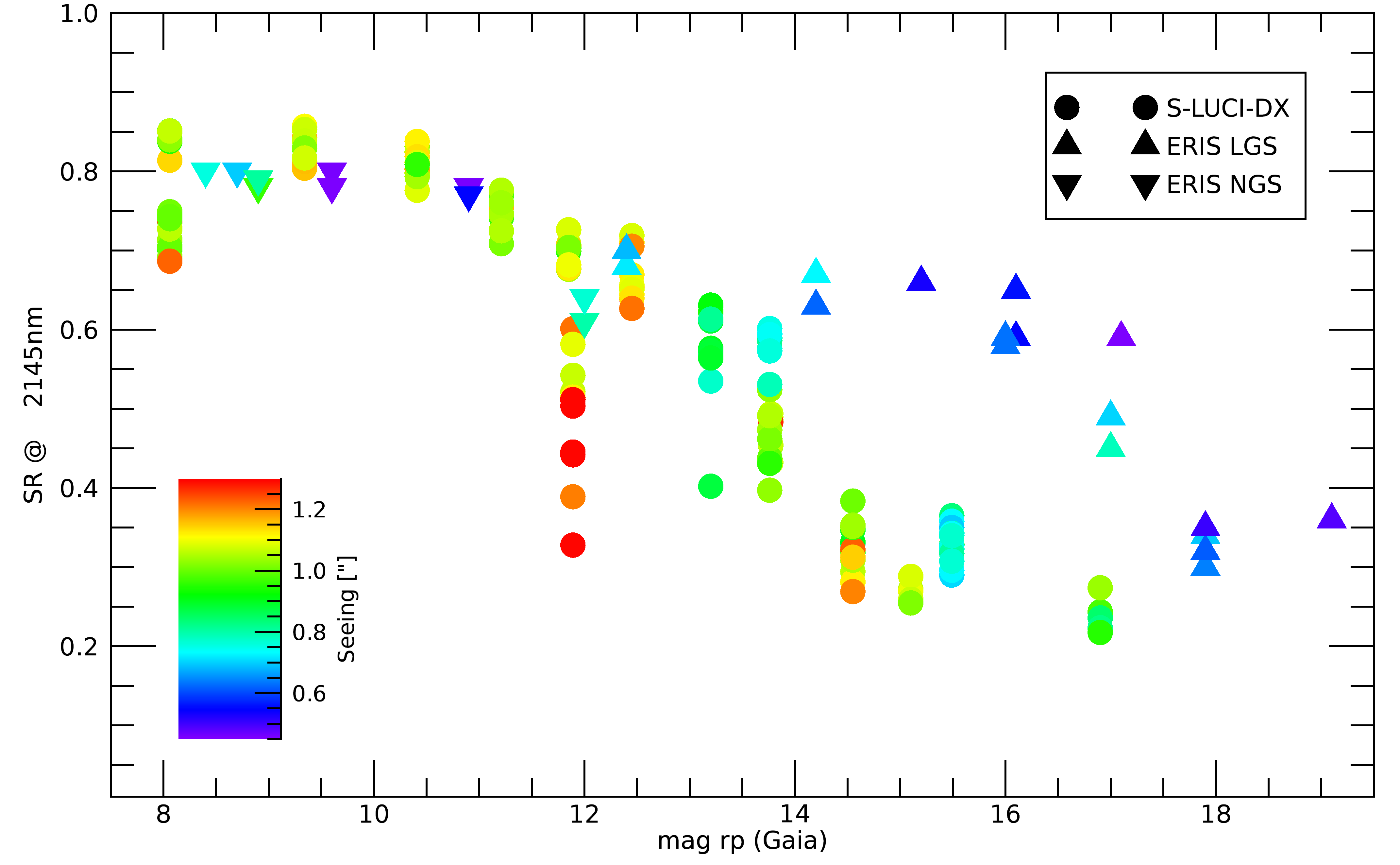}
    \caption{Comparison of SOUL at LBT (circles) performances with those of ERIS at VLT (triangles). The points represent the SR collected on-sky during their respective commissioning. For both systems the SR value is computed on-axis from AO telemetry data. The color represents the seeing along the line of sight, saturated at $1.3$asec. The horizontal axis report the AO reference star magnitude ($G_{RP}$).}
    \label{fig:SOUL_ERIS}
\end{figure}

In order to include the measured performance in a broader context, we compared the SOUL results with those of ERIS-VLT, published this year \cite{ERIS_A&A} at the end of its commissioning.
It is the comparison between two ``general purpose'' SCAO systems, both having an adaptive secondary mirror as corrector. 
The two main difference between the systems are the type of WFS and the availability of a Na Laser Guide Star (LGS). ERIS has 2 Shack-Hartmann WFSs: one to be used with NGS and a second one for the LGS. SOUL is working with a Pyramid WFS with NGS only.
In  fig.~\ref{fig:SOUL_ERIS}, we report the SR values from the ERIS paper \cite{ERIS_A&A} and those from fig.~\ref{fig:SR_SKY} on the right (values from AO telemetry, scaled at 2145nm), as function of the $G_{RP}$ mag.  
We have to take into account that the SOUL data are collected under average worse seeing conditions (color scale) with respect the ERIS ones: the mean value of the seeing is $0.6"$ for ERIS and $1.0"$ for SOUL.
On the bright end ($G_{RP}<10$), we can see that SOUL exceeds ERIS with higher SR value, despite the higher number of actuators on its adaptive secondary mirror (1170 VLT Vs. 672 LBT).
This can be justified with the better rejection of aliasing provided by the PWFS. In the faint range of ERIS NGS ($10<G_{RP}<12$), SOUL keeps performing better, considering the seeing conditions. This is probably due to higher sensitivity of PWFS, providing an higher SNR with the same star flux.
For $G_{RP}>12$ ERIS employs the LGS providing constant performance up to  $G_{RP}>17$. Up to $G_{RP}12.5$ SOUL continues exceeding ERIS, thanks to the possibility to correct 250 modes exploiting the NGS flux with a reduced number of sub-apertures obtained binning on chip $2\times2$. 
For $G_{RP}>14$ SOUL can't compete anymore with ERIS LGS performance. However, SOUL keeps providing $SR(K) >20\%$ for $G_{RP}<17$.

\begin{figure}[ht]
    \centering
    \includegraphics[width=0.65\linewidth]{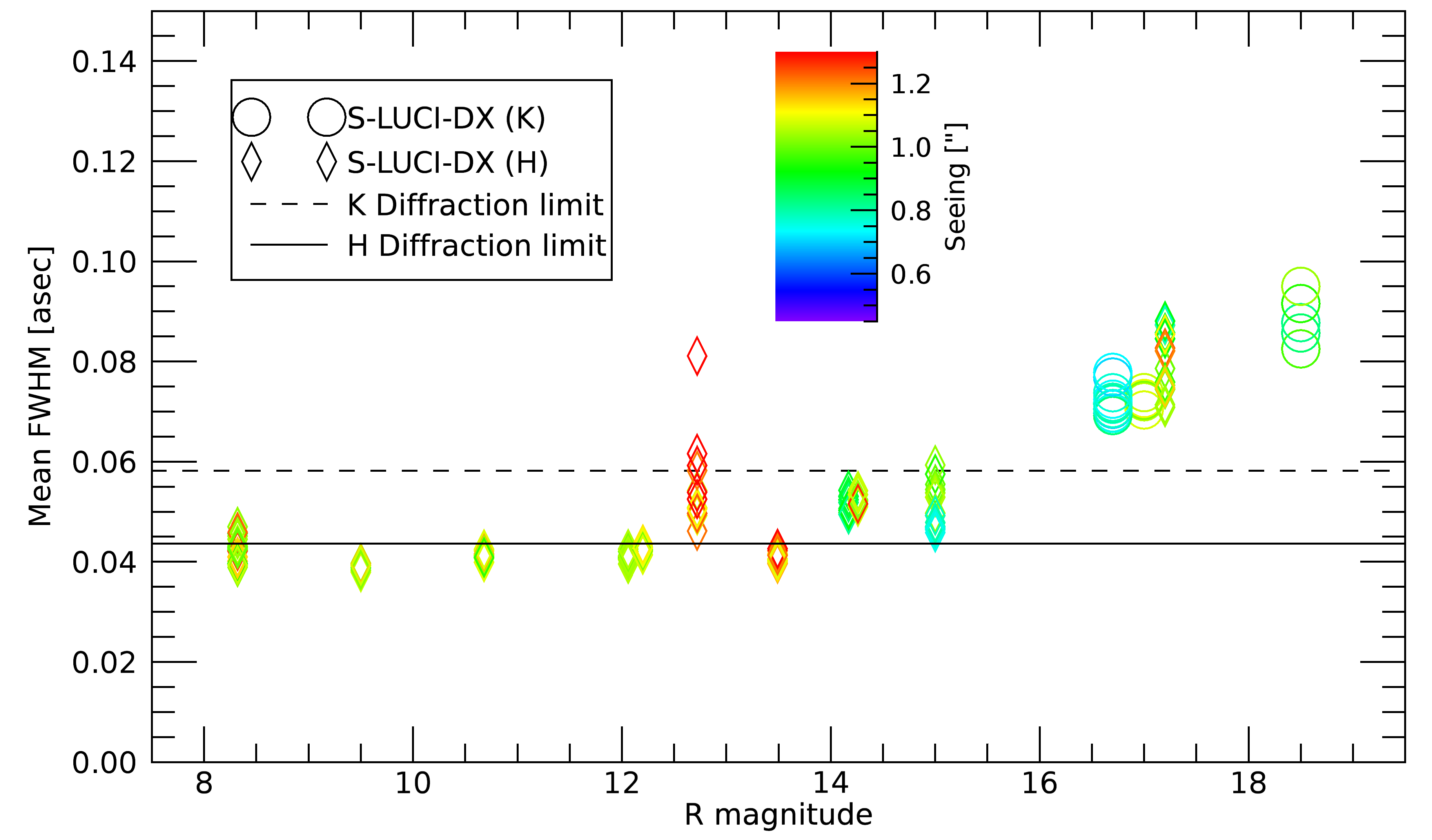}
    \caption{The FWHM of the AO reference image on LUCI2-N30 camera as function of its brightness (catalogue $R$ magnitude). Rhombuses refer to data in H band and FeII, while circles to K band. The lines show the FWHM of the LUCI2-N30 diffraction limit at the the considered wavelengths.  The color represents the seeing along the line of sight, saturated at $1.3$asec.}
    \label{fig:FWHM_Rmag}
\end{figure}

This is one of the key results of the SOUL upgrade.
In fig.~\ref{fig:FWHM_Rmag}, we see the PSF FWHM measured on LUCI-N30 camera as a function of the reference magnitude.
The FWHM is kept below $100mas$ even in the fainter range and for seeing up to $1.0"$. This enables the possibility to use, as AO reference, extra-galactic sources that typically have $R>15$, providing high spatial resolution images. 
Considering the absence of Na LGS facilities at LBT, this results in a key improvement in LBT's offering for the astronomical community.

\section{Early science and next future}
\label{sec:science}

During and in-parallel with the commissioning activity, the SOUL systems have been used for scientific observations. In particular, due to the commissioning sequence (see sect.~\ref{sec:intro}), the most used systems are the 2 on the SX of the telescope.

\begin{figure}[ht]
    \centering
    \includegraphics[width=0.73\linewidth]{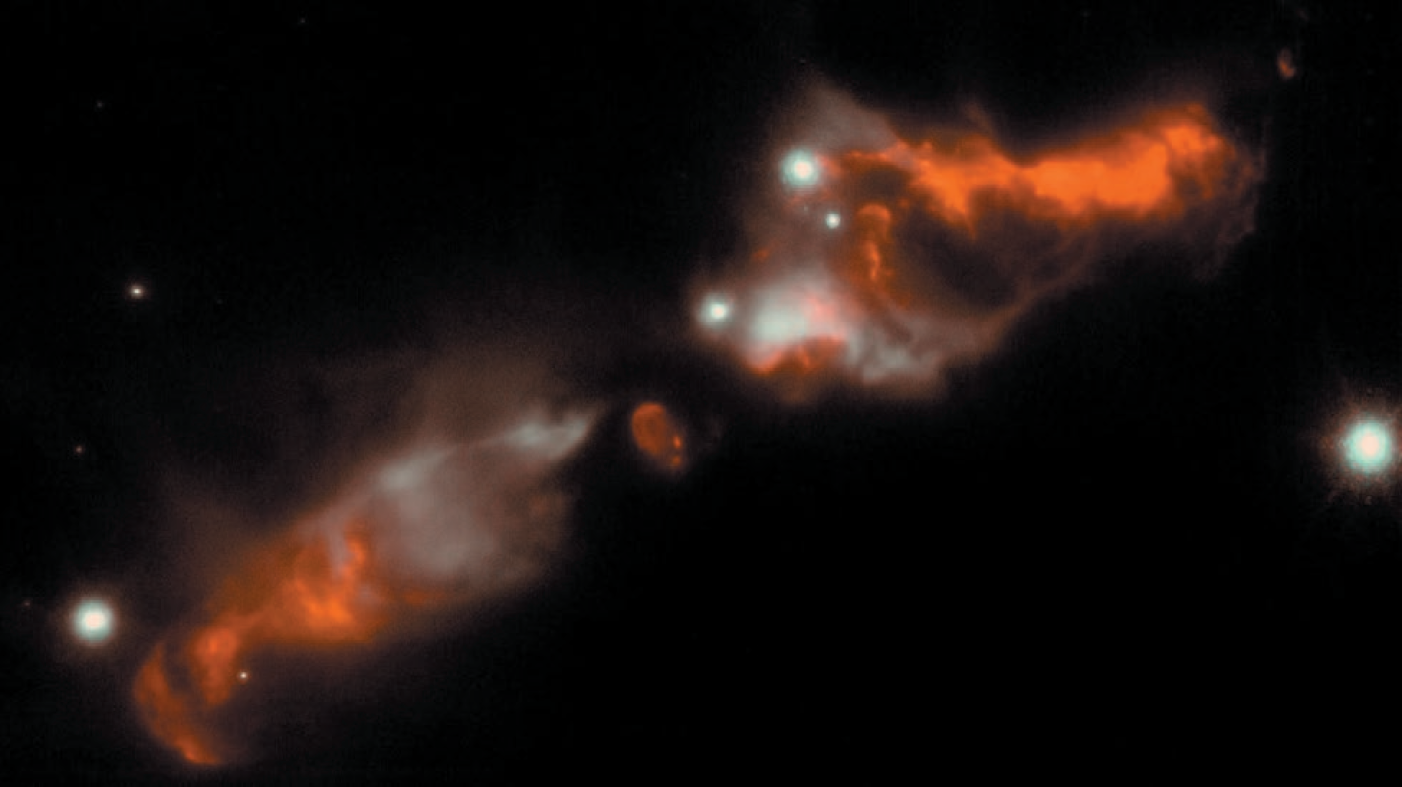}
    \includegraphics[width=0.85\linewidth]{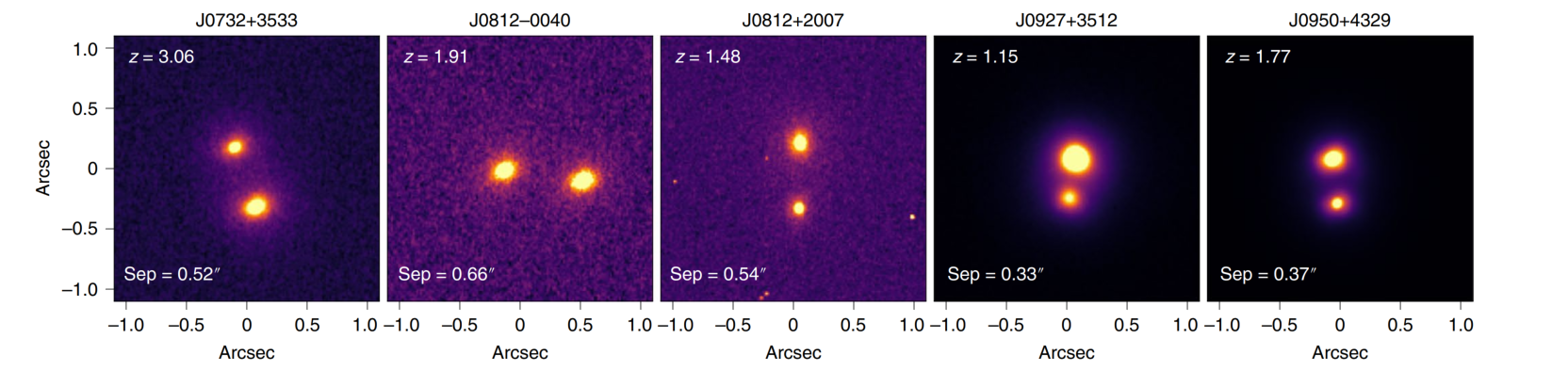}
    \caption{Two example of scientific observations performed with SOUL-LUCI-SX.
    Top: high-mass protostar (IRAS20126 + 4104) in which the ejection of two gas jets in the accretion process is observed \cite{Massi+2023.113M}. The red component of the image corresponds to the emission of molecular hydrogen ($2.12\mu m$), which highlights the shock arcs in both jets, whereas the white component displays the $2.2 \mu m$ thermal radiation of field stars emitted by the protostar and diffused by the dust.
    The star on the right edge is the AO reference $R\sim14$, with a distance from the left edge of th jet of $20$asec.
    Bottom: 5 targets identified as multiple peaks on GAIA catalogues have been confirmed to be double AGNs thanks to SOUL-LUCI-SX \cite{Mannucci+22}. The second of the left has been observed using the AGN  itself ($R=15.3$) as reference for the AO. }
    \label{fig:Sci_images}
\end{figure}

SOUL-LUCI-SX had its very first scientific observation in 2019 \cite{Gilli+19}, followed by, with November 2020, by a set of few nights of "Commissioning Science Time" devoted to the execution of short scientific programs collected in the community of the LBT partners.
These scientific observations, produced useful data for the observers, giving at the same time key feedbacks to the SOUL team for the system tuning. 
Moreover, other observations has been executed in the LBT partners' time. 
In tab.~\ref{tab:SCI_Papers}, we summarize the papers published with SOUL-LUCI-SX data, while others are currently in submissions.
The aim of the table is to provide an outlook on the produced astrophysical science.
In particular, we want to focus your attention toward the presence of both galactic and extra-galactic topics. 
In fig.~\ref{fig:Sci_images} we show two examples of the images produced with SOUL-LUCI-SX.
Other publications, about 10, have been produced with the SOUL-LBTI systems and they are reported in this conference \cite{2023_Ertel_LBTI_AO4ELT7}.

\begin{table}[]
    \centering
    \begin{tabular}{|c|c|c|}
          \textbf{Publication} & \textbf{Topic}  & \textbf{Reference} \\
          \hline
          Gilli+19 A\&A & Galaxy cluster& \cite{Gilli+19} \\
          Zurlo+22 A\&A & Exoplanets    & \cite{Zurlo+22}  \\
          Rossi+22 ApJ & Supernova  & \cite{Rossi+22}\\
          Mannucci+22 NatAstr & Double AGN & \cite{Mannucci+22}\\
          Annibali+23 ApJL & Extrag. stellar pop. & \cite{Annibali+23} \\
          Massi+23 A\&A & Star formation & \cite{Massi+2023.113M}\\
          Fedriani+23 A\&A & Star formation &   \cite{Fedriani+23}\\
    \end{tabular}
    \caption{The table report the list of published astrophysical results obtained with SOUL-LUCI-SX together with their class of the science topic and  reference.}
    \label{tab:SCI_Papers}
\end{table}

The next steps for SOUL systems involves multiple aspects; we mention here a selection of the most significant ones.

A relevant step forward in time efficiency and performance stability can be done implementing the "Modal Gain Machine", as proposed and tested on-sky in 2019 \cite{Agapito_Control_MNRAS2021}. This is an automatic modal gain optimization updated on the fly during observations.
This will provide, at the same time, a better optimization of the loop, together with a faster and more robust AO bootstrap. 

Regarding providing the support to the observer's data reduction, we know how valuable a PSF model is. There are two promising technique that we would like to include in a future tool for observers. First, on-axis PSF reconstruction based on AO telemetry data \cite{2022_Simioni_PSF-SOUL_JATIS} \cite{2022_Arcidiacono_BRUTE_PSF_SOUL_SPIE}. Second, using fast numerical tools, like TIPTOP \cite{2020Neichel_TIPTOP_SPIE}, to provide off-axis PSF model, based on the given observational conditions.  

Finally, the biggest step forward are the second generation LBT instruments  that are aimed to exploit SOUL at its best performance.
The new instruments are: SHARK-NIR \cite{2022Farinato_SHARKNIR_SPIE}, SHARK-VIS \cite{2022Pedichini_SHARK-VIS_SPIE} and iLocater \cite{2023Crass_iLocater_AAS}.
The first is a NIR imager and coronagraph, optimized for high contrast and it is currently completing its commissioning at the LBTI-SX focal stations.
SHARK-VIS is the analogous for shorter wavelengths down to $400$nm and it is ready for the installation at the LBTI-DX focal station.
iLocator is an extremely precise radial velocity (EPRV) spectrograph in the NIR fed by both SOUL-LBTI focal stations.
The capabilities of these instruments surely represent a new challenge for the SOUL systems, requiring dedicated tuning in order to exploit them at their best.

\section{Conclusions}
\label{sec:conclusions}
The SOUL project completed its path upgrading the 4 LBT's SCAO systems that are now offered for science operations to the astronomical community. 
The work done in the project allowed us to obtain the expected performances in all the magnitude range. 
The ability to provide high spatial resolution down to $G_{RP}=17$ is ground breaking for NGS SCAO systems, with remarkable consequences to offer extragalactic science at LBT.
A relevant part of the commissioning work has been devoted to improve the robustness and operability of the systems. The astrophysical results already published confirm good results on these aspects too. 
We hope that the SOUL's experience can be beneficial for the AO community engaged today in the design of SCAO systems for the 25-40m generation of telescopes both in terms of achievable performances and control strategy to deal with pyramid WFS. 
The SOUL team is now looking forward to the next generation of LBT instruments in order to exploit all SOUL's capabilities.

\acknowledgments 

The authors want to thank the LBT mountain crew and all LBTO personnel for the hard work supporting the long commissioning activity of the SOUL systems.

Observations with SOUL systems have benefited from the use of ALTA Center (alta.arcetri.inaf.it) forecasts performed with the Astro-Meso-Nh model.
Initialization data of the ALTA automatic forecast system come from the General Circulation Model (HRES) of the European Centre for Medium Range Weather Forecasts.

\printbibliography 

\end{document}